\documentclass[sigconf]{acmart}

\usepackage{booktabs}
\usepackage{amsmath}
\usepackage{array}
\usepackage{graphicx}
\usepackage{multirow}

\setcopyright{none}
\copyrightyear{2026}
\acmYear{2026}
\acmDOI{}
\acmConference[AgenticDev '26]{International Workshop on Agentic AI for Next-Generation Software Development}{October 12, 2026}{Munich, Germany}
\acmISBN{}

\ccsdesc[500]{Software and its engineering~Software testing and debugging}
\ccsdesc[500]{Computing methodologies~Natural language generation}
\ccsdesc[300]{Computing methodologies~Planning and scheduling}

\keywords{agentic AI, tool-use agents, synthetic traces, execution-grounded data, supervised fine-tuning}

\title[Trace Synthesis for LLM Agents]{Execution-First Synthetic Tool-Use Trace Generation for LLM Agents}

\author{Hafsa Ouajdi}
\authornote{Both authors contributed equally to this research.}
\affiliation{%
  \institution{EURECOM}
  \city{Sophia Antipolis}
  \country{France}}
\affiliation{%
  \institution{Aily Labs}
  \city{Munich}
  \country{Germany}}
\email{hafsa.ouajdi@ailylabs.com}

\author{Francesco Giannuzzo}
\authornotemark[1]
\affiliation{%
  \institution{EURECOM}
  \city{Sophia Antipolis}
  \country{France}}
\affiliation{%
  \institution{Aily Labs}
  \city{Munich}
  \country{Germany}}
\email{Francesco.Giannuzzo@eurecom.fr}

\author{Alaa Boukhary}
\affiliation{%
  \institution{Aily Labs}
  \city{Munich}
  \country{Germany}}
\email{alaa.boukhary@ailylabs.com}

\author{Paolo Papotti}
\affiliation{%
  \institution{EURECOM}
  \city{Sophia Antipolis}
  \country{France}}
\email{papotti@eurecom.fr}

\author{Gerard Conangla}
\affiliation{%
  \institution{Aily Labs}
  \city{Munich}
  \country{Germany}}
\email{gerard.conangla@ailylabs.com}

\author{Adam Elwood}
\affiliation{%
  \institution{Aily Labs}
  \city{Munich}
  \country{Germany}}
\email{adam.elwood@ailylabs.com}

\begin{abstract}
Agentic software-engineering and industrial systems increasingly operate through executable workflows rather than code generation alone: they search artifacts, invoke tools, inspect structured observations, and query databases. Training these agents requires supervision data that captures valid tool interactions and executable workflows. However, traditional query-first data synthesis can fail because plausible user requests may not correspond to valid tool sequences, compatible parameters, or available data. To address this limitation, we propose \textsc{SyntheticAgentTraceQA}, an execution-first framework for generating scalable supervision data for tool-augmented agents. Our framework first constructs high-level workflow structures, maps them to available tools through dependency-aware assignment, executes and validates the resulting traces in controlled environments, and only then synthesizes natural-language user tasks, teacher-generated reasoning annotations, and reference answers. We evaluate the framework across four tool ecosystems and use the resulting data to fine-tune and evaluate Qwen model variants. The results show that execution-grounded supervision improves tool execution behavior, reference-trace agreement, and answer-generation performance on the evaluated tasks. Further analysis reveals a supervision trade-off: masked supervision, which excludes reasoning annotations from the training objective, improves final-answer metrics, whereas full supervision, computing loss over the complete assistant output including reasoning tokens, underperforms on answer quality and does not consistently improve reference-trace agreement, particularly at the 9B scale. These findings highlight the importance of designing synthetic supervision according to the desired capabilities of tool-augmented agents.
\end{abstract}

\begin{document}
\maketitle
\section{Introduction}

Large language models (LLMs) have enabled a new generation of autonomous agents capable of interpreting complex user requests and interacting with external environments. A particularly effective paradigm is that of \textbf{\textit{tool-augmented agents}}, which extend the capabilities of LLMs by invoking external tools such as APIs, data-analysis engines, search systems, and other specialized services~\cite{yao2023react,schick2023toolformer}. By integrating these resources, LLM-based agents move beyond next-token prediction toward dynamic, interactive problem solving~\cite{wang2024survey,xi2025rise}.

Despite these advances, these agents can still fail in complex, multi-step technical workflows. Common failure modes include selecting inappropriate tools~\cite{yao2024taubench,lu2025toolsandbox}, generating invalid tool arguments (e.g., incorrect parameter names or data types), failing to incorporate intermediate execution results into subsequent reasoning, and producing responses that are not fully supported by the evidence gathered during execution~\cite{soni2026toolfailbench,englander2026agents,lu2025toolsandbox}. These failures become particularly consequential in software engineering tasks involving artifact retrieval, programmatic tool execution, structured data access, and the orchestration of multiple API calls, where correctness depends on accurate tool use and faithful reasoning over intermediate observations~\cite{yang2024swe,jimenez2024swe,lu2025toolsandbox,sahoo2026agentlens}. They also compound broader usability and oversight challenges in deployed agent systems~\cite{Shome_2026}.

This paper focuses on a specific aspect of agent performance rather than general coding ability. Specifically, we study whether agents can construct and execute multi-step tool workflows whose correctness can be objectively verified through tool outputs and environment states. Across all tasks, the central capability under evaluation is reliable reasoning over tools, schemas, and intermediate execution results.

Existing benchmarks and data-synthesis methods often rely on manually crafted~\cite{guo2024stabletoolbench,qin2023toolllm} or forward-generated~\cite{wang2023selfinstruct,li2023toolalpaca,zhou-etal-2026-toolgrad} scenarios, which are costly, difficult to scale, and may not guarantee alignment between user requests and executable tool traces. In a conventional query-first pipeline, a plausible user request may require unavailable tools, invalid parameters, or execution paths that cannot actually solve the task~\cite{toolbehonest}.

To address this limitation, we introduce \textbf{SyntheticAgentTraceQA}, a backward-chaining data-generation pipeline that retains only traces that execute successfully in the generation environment and uses them as supervision signals. Rather than starting from a user request, our approach first constructs and validates executable tool traces in a controlled environment before synthesizing natural-language tasks. This execution-first strategy filters or reduces common synthesis errors, such as tool-capability mismatches and invalid parameterizations, while producing training data grounded in executable workflows.

\paragraph{Research Questions.}
\begin{itemize}
\setlength{\itemsep}{0pt}
\setlength{\parsep}{0pt}
\setlength{\topsep}{0pt}
\setlength{\partopsep}{0pt}

\item \textbf{(RQ1)} Does execution-grounded fine-tuning improve tool behavior, reference-trace agreement, and answer-generation performance?

\item \textbf{(RQ2)} How does model scale influence the effects of execution-grounded supervision across these evaluation metrics?

\item \textbf{(RQ3)} How does supervising reasoning tokens affect tool behavior, reference-trace agreement, and answer-generation performance?
\end{itemize}

\paragraph{Contributions.}
\begin{itemize}
\setlength{\itemsep}{0pt}
\setlength{\parsep}{0pt}
\setlength{\topsep}{0pt}
\setlength{\partopsep}{0pt}
\item \textbf{The SyntheticAgentTraceQA Pipeline:} An execution-first framework for generating synthetic supervision data for tool-augmented agents through validated execution traces with varying levels of workflow and task complexity.

\item \textbf{Multi-Domain Taxonomy \& Extension:} An operational taxonomy for tools and parameters that unifies heterogeneous tool environments—including internal company tools (finance, research portfolio, supply chain) and the \texttt{genius\_song\_lyrics} tool group from ToolBench for music search and metadata retrieval—under a single abstract template and profiling framework. Tool execution is performed through local wrappers around the ToolBench APIs using cached responses.

\item \textbf{Controlled Evaluation:} A fine-tuning study showing measured improvements in tool-use reliability, Reference-Trace Agreement, and answer quality across multiple model variants.

\end{itemize}
\section{Related Work}
\paragraph{Tool-augmented language models.}
Large language models (LLMs) have demonstrated strong performance across a wide range of tasks. However, their capabilities remain limited by the static knowledge acquired during pretraining. Tool augmentation overcomes this limitation by enabling access to external resources such as search engines, databases, computational systems, and APIs. This extends LLMs with up-to-date information and specialized functionalities~\cite{komeili2022internet,gou2024critic,gu2024middleware,huang2024metatool,qu2025tool}. In addition, tools allow LLMs to interact with external environments, supporting task automation and execution beyond text generation~\cite{zhuang2023toolqa,qin2023toolllm}. Tool use can also improve response grounding, transparency, and interpretability by exposing intermediate reasoning and evidence sources~\cite{armengol2025execute}.

Several approaches have been proposed to facilitate tool use in language models. Toolformer introduced a self-supervised framework for learning API invocation behavior from data~\cite{schick2023toolformer}, while ReAct demonstrated the effectiveness of interleaving reasoning and actions during task execution~\cite{yao2023react}. These paradigms have subsequently influenced a broad range of tool-augmented and agent-based systems~\cite{qin2023toolllm,liu2024toolnet}.

Although these approaches significantly improve the capabilities of language agents~\cite{amugongo2025retrieval,Bran2023Augmenting,Goodell2025Large}, they primarily focus on inference-time tool use. Comparatively less attention has been given to using tool interactions as supervision signals, particularly through execution-grounded traces for training and evaluating tool-augmented agents.
\paragraph{Structured reasoning and planning.}

Recent work has increasingly emphasized process-centric reasoning over outcome-only evaluation. Approaches such as Chain-of-Thought (CoT) and Tree of Thoughts (ToT) demonstrate that exposing intermediate reasoning steps and exploring alternative trajectories can improve problem solving by enabling decomposition, self-verification, and search~\cite{wei2022chain,yao2023treethoughtsdeliberateproblem}. Building on this idea, reasoning traces have become an important source of supervision for training and evaluating reasoning-oriented models. Evidence suggests that structured trajectories capture procedural knowledge beyond final answers alone~\cite{xie2025agentsynth,luo2026ise}.

Recent work has also explored synthetic trace generation and trajectory modeling, where traces explicitly represent goals, plans, tool invocations, observations, and recovery actions. Frameworks such as AgentSynth and ISE generate execution-grounded trajectories for analyzing and supervising complex reasoning behavior~\cite{xie2025agentsynth,luo2026ise}. These works support our focus on execution-grounded tool-use traces as a scalable source of supervision and evaluation beyond final-task correctness.

\paragraph{Synthetic data and instruction tuning.}

Most instruction-tuning pipelines follow a forward-generation paradigm: an LLM is prompted with a small seed set to generate new instructions, inputs, and outputs, which are subsequently filtered and used for supervised fine-tuning~\cite{wang2023selfinstruct,Wang2024A}. Approaches such as Self-Instruct, Alpaca, and GPT-4-based distillation differ in the teacher model and prompting strategy but largely preserve the same instruction $\rightarrow$ input $\rightarrow$ output generation process~\cite{wang2023selfinstruct,Wang2023Harnessing,Peng2023Instruction,Zhang2023Instruction}. Tool-use data generation extends this paradigm by conditioning on tool or API descriptions and synthesizing tool-using tasks together with candidate call sequences or solution paths~\cite{qin2023toolllm,zhang2024geospatial,xu2025toucan}. While effective for scaling supervision, these approaches often rely on static prompts and model-generated trajectories, which can introduce noise, unrealistic behaviors, and invalid multi-tool interactions~\cite{qin2023toolllm,huang2026fintoolsyn,koksal2024muri}.

Execution-first approaches reverse this pipeline by treating valid executions or trajectories as the primary artifact. Rather than generating a task and inferring a solution path, they begin with verified interactions and derive traces or user-facing tasks from successful executions~\cite{armengol2025execute,wang2026trajectory2task}. Recent trajectory-centric methods share a similar motivation but often rely on proxy signals instead of verified executions. ToolGrad generates and refines candidate workflows using textual feedback, which may increase workflow complexity without guaranteeing correctness~\cite{zhou-etal-2026-toolgrad}. ToolMind constructs trajectories through similarity-based tool graphs and random walks, which can produce unrealistic or non-executable workflows that do not reflect real user goals~\cite{yang2025toolmind}. HardGen leverages API dependency graphs to generate challenging training examples, but its effectiveness depends on accurate dependency metadata and predefined tool relationships~\cite{hao2026failure}.
\paragraph{Agent benchmarks.}

Recent benchmarks for tool-augmented agents have evolved from broad API-centric evaluations toward more reproducible and production-oriented execution environments. Early efforts such as ToolLLM and API-Bank were built around diverse real-world APIs and evaluated planning, retrieval, and tool invocation through executable interactions~\cite{qin2023toolllm,li2023api}. Subsequent work identified important reproducibility challenges, showing that benchmark performance can be affected by API drift, unstable tool availability, and complex multi-turn interactions~\cite{guo2024stabletoolbench,yao2024taubench,dong2025toolplaygrounds}. More recent benchmarks address these issues through virtualized APIs, simulated tool ecosystems, and controlled execution environments that better reflect long-horizon interactions and realistic user workflows~\cite{guo2024stabletoolbench,yang2026abcbench,shen2026tripbench,Yu2026Benchmarking}.

A complementary line of research investigates synthetic data generation for tool-augmented agents. Recent work has shown that producing high-quality supervision often requires generating a large pool of candidate examples before selecting those that satisfy the desired quality criteria. For example, Autodata and Toucan~\cite{kulikov2026autodata,xu2025toucan} employ an iterative framework in which candidate examples are repeatedly generated, evaluated, and filtered through multiple validation stages. While this process can improve dataset quality, it also increases generation cost by requiring substantial oversampling before selection.

Collectively, these efforts highlight the importance of executable supervision for tool-augmented agents. Existing approaches either focus on evaluating agent behavior or generate execution traces from predefined dependencies or proxy supervision. In contrast, \textsc{SyntheticAgentTraceQA} combines an operational tool and parameter taxonomy, active-domain profiling, LLM-generated abstract execution DAGs, dependency-aware tool assignment, execution-driven validation before task synthesis, and application across four heterogeneous tool ecosystems within a single generation framework. This combination enables scalable generation of execution-grounded supervision while improving sample efficiency through early validation of executable traces.
\section{Problem Setup \& Formalization}

Let $T = \{T_1, \dots, T_n\}$ be a set of available tools. Each tool $T_i$ maps a set of input arguments $x_i$ to an observation output $y_i$. An execution trace $\tau$ is defined as a sequence of tool invocations:
\[
\tau = [(T_1, x_1, y_1), \dots, (T_m, x_m, y_m)].
\]

Given a natural-language user task $z$, a tool-augmented agent must generate a predicted execution trace $\hat{\tau}$ and a final text answer $\hat{r}$ grounded in the accumulated observations. Our objective is to systematically generate high-quality training tuples $(z, \tau, r)$, where $\tau$ is fully executable, $r$ is supported by the execution outputs of $\tau$, and $z$ reflects a realistic user request.

\paragraph{Scope.}
Although our task space extends beyond conventional software engineering benchmarks centered on source code editing and repository bug fixing, it remains fundamentally software-engineering in nature. Tasks require agents to interact with software artifacts such as code repositories, system logs, structured data, domain-specific APIs, database tables, and analytical scripts. Success depends on the same core competencies required for agentic software engineering: composing multi-step tool workflows, respecting interface and schema constraints, executing syntactically valid operations, incorporating execution feedback into subsequent reasoning, and avoiding unsupported or unverifiable responses.

\section{The SyntheticAgentTraceQA Pipeline}
\begin{figure*}[t]
\centering
\includegraphics[width=1\textwidth]{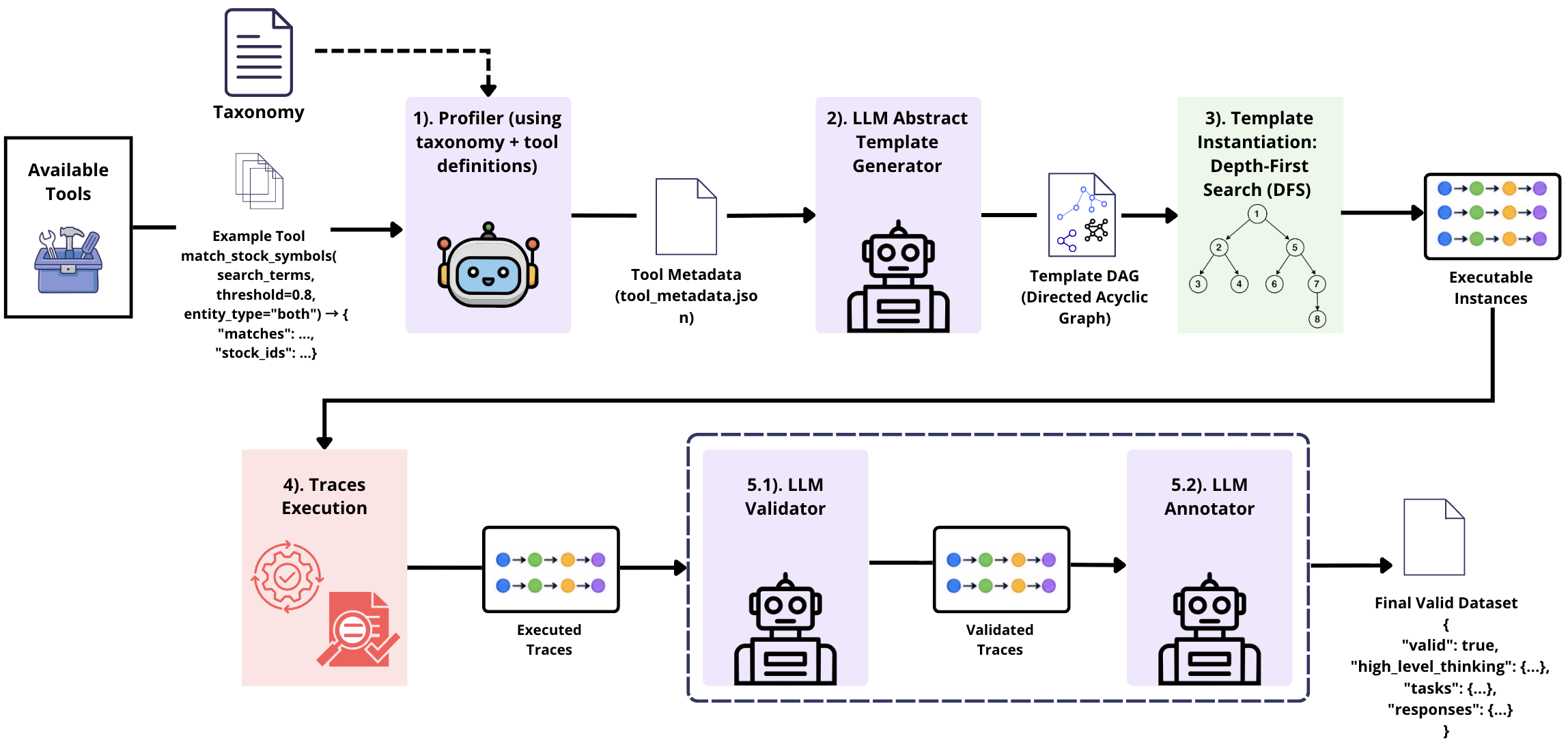}
\caption{Overview of SyntheticAgentTraceQA. The execution-first pipeline constructs and validates tool-use traces before synthesizing user tasks. (1) \textbf{Profiler} extracts tool metadata and parameter roles; (2) \textbf{Template Generator} creates abstract DAG-based workflows; (3) \textbf{DFS} instantiates executable traces under data-flow constraints; (4) \textbf{Trace Execution} validates candidate workflows; and (5) \textbf{Validation \& Task Synthesis} generates user tasks and answers from validated traces.}
\label{fig:pipeline}
\end{figure*}

To overcome the scalability limitations and capability gaps of language-first data generation, we introduce \textbf{SyntheticAgentTraceQA}, an execution-first pipeline. Instead of starting from a natural-language query, our approach first constructs and executes valid tool traces, then synthesizes user tasks only after execution-based validation. Figure~\ref{fig:pipeline} summarizes the five stages of the pipeline: \textit{\textbf{Profiler}} extracts tool metadata; \textit{\textbf{Template Generator}} produces abstract execution templates; \textit{\textbf{Depth-First Search (DFS)}} constructs executable tool traces; \textit{\textbf{Trace Execution}} executes the generated traces and \textit{\textbf{Validation \& Task Synthesis}}, which validates the executed traces and synthesizes corresponding user tasks from the validated executions.

\subsection{Operational Taxonomy}\label{taxonomy}

To bridge the gap between abstract user intent and executable tool invocations, we organize the available tools into a finite set of \emph{Operational Classes}. These classes provide the first level of abstraction in our execution-first framework. Instead of reasoning over individual tools, the pipeline constructs execution templates using functional categories, decoupling high-level planning from tool-specific implementations while preserving the logical dependencies between successive operations.

\[
\mathcal{O} =
\{\textsc{Matcher},
\textsc{Retriever},
\textsc{Analyzer},
\textsc{Aggregator},
\textsc{Metadata},
\textsc{Access}\}.
\]

Each operational class defines a distinct functional role:

\begin{itemize}
\setlength{\itemsep}{0pt}
\item \textbf{Matcher}: Resolves ambiguous user-provided entities into unique system identifiers.
\item \textbf{Retriever}: Retrieves data and attributes associated with resolved entities.
\item \textbf{Analyzer}: Processes retrieved data to derive insights, detect patterns, or generate predictions.
\item \textbf{Aggregator}: Combines information from multiple entities or observations through statistical or logical operations.
\item \textbf{Metadata}: Provides schema information, field definitions, and configuration details.
\item \textbf{Access}: Loads and validates data sources for downstream processing.
\end{itemize}
\subsection{Tool and Data Profiling}

The first stage of our pipeline is the \textit{Profiler}, which constructs an enhanced metadata catalog for each tool. It is implemented using \textit{Claude Sonnet 4.6}; no manual annotation is performed, and the resulting metadata are used during template generation and compatibility checking.
The profiler automatically assigns each tool to an operational class (Matcher, Retriever, Analyzer, Aggregator, Metadata, or Access) and classifies every input parameter and output field according to its functional role:

\begin{itemize}
\setlength{\itemsep}{0pt}
\item \textbf{Control} (\(P_{\mathrm{ctrl}}\)): task-defining inputs that remain consistent throughout an execution trace.
\item \textbf{Data-flow} (\(P_{\mathrm{flow}}\)): inputs supplied by upstream tool outputs.
\item \textbf{Temporal} (\(P_{\mathrm{time}}\)): parameters defining temporal constraints.
\item \textbf{Configuration} (\(P_{\mathrm{set}}\)): tool-specific settings independent of task intent.
\end{itemize}

For a tool \(T_i\),

\[
\mathrm{Params}(T_i)
=
P_{\mathrm{ctrl}}(T_i)
\cup
P_{\mathrm{flow}}(T_i)
\cup
P_{\mathrm{time}}(T_i)
\cup
P_{\mathrm{set}}(T_i).
\]

Each parameter is represented by its name, data type, semantic role, and optionality. The profiler also constructs an \emph{Active Domain}, i.e., a set of feasible values obtained by sampling the underlying data or extracting values from tool specifications.

Formally,

\[
\Phi:
(T_i,\mathrm{Params}(T_i),\mathrm{Schema}(T_i))
\rightarrow
\mathcal{M}^{*}(T_i),
\]

where \(\mathcal{M}^{*}(T_i)\) augments the tool schema with parameter roles, active domains, and explicit compatibility metadata used during template instantiation. Compatibility between tool calls is determined by semantic roles, data types, and profiler-discovered source mappings rather than parameter names alone.

\subsection{Abstract Template Generation}

Rather than generating tool calls directly from a user query, our execution-first pipeline first constructs an abstract execution template  \(A_i\) as a Directed Acyclic Graph (DAG), \(G_{A_i}\), whose nodes represent abstract operational blocks rather than concrete tools.

Formally, let
\(\mathcal{O}\) denote the set of operational classes defined in Section~\ref{taxonomy},
\(\mathcal{M}^{*}\) the enhanced metadata catalog produced by the Profiler,
and \(\mathcal{A}\) the space of abstract templates.
An abstract template \(A_i\in\mathcal{A}\) is represented as

\[
G_{A_i}=(V_{A_i},E_{A_i}),
\]

where

\begin{itemize}
\setlength{\itemsep}{0pt}
\item \(V_{A_i}=\{b_1,\ldots,b_L\}\) is the set of operational blocks;
\item \(E_{A_i}\subseteq V_{A_i}\times V_{A_i}\) is the set of directed execution and data-flow dependencies.
\end{itemize}

Each operational block \(b_l\in V_{A_i}\) is defined as

\[
b_l=
(o_l,\mathrm{pool}_l,\mathrm{dep}_l,\mathrm{desc}_l),
\]

where \(o_l\in\mathcal{O}\) denotes the operational class,
\(\mathrm{pool}_l\subseteq\mathcal{T}\) is the set of compatible tools,
\[
\mathrm{dep}_l=\{\,b_r\mid(b_r,b_l)\in E_{A_i}\,\},
\]
is the set of predecessor blocks, and
\(\mathrm{desc}_l\) is a domain-specific description used during task synthesis.

To encourage structural diversity, templates are generated at three complexity levels according to the number of operational blocks:

\[
\mathcal{D}=
\begin{cases}
\textsc{Easy}, & |V_{A_i}|\in\{1,2\},\\
\textsc{Medium}, & |V_{A_i}|\in\{3,4\},\\
\textsc{Hard}, & |V_{A_i}|\in\{5,6\}.
\end{cases}
\]

Template generation is performed by an LLM,

\[
G:\mathcal{O}\times\mathcal{M}^{*}\times\mathcal{D}\rightarrow\mathcal{A},
\]

where \(\mathcal{D}\) specifies the target template complexity. The generated templates satisfy three constraints: (i) data-flow consistency, ensuring required inputs are provided by predecessor blocks; (ii) tool interchangeability, associating each operational block with a pool of compatible tools; and (iii) complexity control, constraining the DAG topology to the selected complexity level.
\subsection{Template Instantiation}

Given an abstract template \(A_i\), this stage performs \emph{tool selection}: each abstract operational block is replaced with a concrete tool. Since multiple tools may satisfy the same operational class, instantiation produces a set of candidate execution traces:

\[
\mathcal{I}_{A_i}
=
\{I_{c,1},I_{c,2},\ldots,I_{c,n}\}.
\]

For example, consider an abstract template of three operational blocks, \emph{Matcher} $\rightarrow$ \emph{Retriever} $\rightarrow$ \emph{Analyzer}. Instantiation replaces these abstract operations with concrete tools, e.g., \texttt{match\_company} $\rightarrow$ \texttt{retrieve\_financial\_statement} $\rightarrow$ \texttt{calculate\_growth}. At this stage only the tool \emph{identities} are fixed; the concrete argument values they operate on (e.g., a specific company and fiscal year) are not chosen here but are bound later, during trace execution (Section~\ref{subsec:trace_execution}).

Because the number of possible assignments grows combinatorially, we employ a Depth-First Search (DFS) strategy to explore valid tool combinations efficiently while pruning incompatible assignments. The procedure consists of three steps:

\begin{itemize}
\setlength{\itemsep}{0pt}

\item \textbf{Sequential processing}: The template DAG is traversed in topological order. For each operational block \(b_l\), the search iterates over the candidate tools in \(\mathrm{pool}_l\). An instantiation is complete once every block has been assigned a compatible tool.

\item \textbf{Compatibility checking}: For each candidate tool \(T_i\in\mathrm{pool}_l\), the algorithm verifies that all required input fields are available from predecessor outputs. Formally,

\[
P_{\mathrm{flow}}(T_i)\subseteq V_{\mathrm{visible}},
\]

where \(V_{\mathrm{visible}}\) contains the outputs of previously instantiated tools. Compatibility is determined using the profiler metadata: an upstream output is considered compatible with a downstream input when their semantic roles, data types, and explicit source mappings are consistent. Parameter names alone are not used to establish compatibility.
\item \textbf{State propagation}: The visible state \(V_{\mathrm{visible}}\) is the set of data fields currently available from all previously instantiated tools. After selecting a compatible tool \(T_i\), its output fields are added to the visible state,

\[
V_{\mathrm{visible}}
\leftarrow
V_{\mathrm{visible}}
\cup
\mathrm{OutputFields}(T_i),
\]

making them available as inputs for subsequent tools. The search then proceeds recursively to the next operational block. Whenever no compatible assignment exists, the DFS backtracks and explores an alternative branch.

\end{itemize}

\subsection{Trace Execution}
\label{subsec:trace_execution}

Each candidate instance is executed to verify that the abstract plan is feasible in the target data environment. Whereas instantiation determines \emph{which} tools are selected, execution determines \emph{which values} they are invoked with and runs them. Execution maintains a context

\[
C=(C_{\mathrm{in}},C_{\mathrm{out}},W),
\]

where \(C_{\mathrm{in}}\) stores bound input parameters, \(C_{\mathrm{out}}\) stores tool outputs for downstream use, and \(W\) maintains temporal consistency throughout the workflow.

Execution proceeds in topological order. At each step, the required arguments for tool \(T_k\) are resolved from the current context or sampled when unavailable, producing \(\mathrm{bind}(x_k)\). The tool is then executed, yielding output \(y_k\), and the context is updated as

\[
C^{(k+1)}=
\left(
C_{\mathrm{in}}^{(k)}\cup\mathrm{bind}(x_k),
\;
C_{\mathrm{out}}^{(k)}\cup\{(T_k,y_k)\},
\;
W^{(k)}\cup W_k
\right).
\]

A candidate is retained only if all tool calls pass runtime validation, including tool availability, schema-compliant arguments, successful execution, usable outputs, and satisfied dependencies; otherwise it is discarded. To improve robustness, the execution engine re-samples non-deterministic parameters and adaptively relaxes matcher thresholds when appropriate.

\subsection{Validation \& Task Synthesis}

The final stage transforms a successfully executed trace \(I_e\) into a validated dataset instance \(I_v\) using two LLM-based components: a \emph{Validator}, which verifies the executed trace, and an \emph{Annotator}, which generates the final dataset entry.

\paragraph{Validation.}
Crucially, because our pipeline generates data in reverse, \emph{no user question exists at this point}. Validation therefore assesses only the consistency of the executed trace, rather than its agreement with a user query. Specifically, it verifies the structural consistency of the execution graph, correct parameter propagation across tool calls, compliance with tool schemas, temporal consistency, and the overall semantic coherence of the workflow. Traces that fail any validation criterion are discarded. The user question, reasoning trace, and reference answer are synthesized only after a trace has successfully passed validation. This validation is performed using the same LLM used to generate the trace.
\paragraph{Task synthesis.}
For each validated execution trace, an LLM generates three user tasks (\textit{Basic}, \textit{Intermediate}, and \textit{Challenging}), together with the corresponding teacher-generated reasoning annotations and reference answers. These difficulty levels characterize the formulation of the user request rather than the complexity of the underlying execution trace. The \textit{Basic} task explicitly specifies the operational intent and required entities, the \textit{Intermediate} task omits some execution details that must be inferred from context, and the \textit{Challenging} task expresses a high-level or under-specified objective requiring greater interpretation, while all three correspond to the same validated execution trace. This design exposes the same executable workflow through multiple natural-language formulations, increasing linguistic and behavioral diversity while preserving the underlying task intent and execution semantics.

\section{Multi-Domain Extensions \& Tool Sets}
SyntheticAgentTraceQA is designed to be domain-agnostic: the same execution-first pipeline operates across environments with different schemas and tool sets without modifying the generation process. We evaluate the framework on four domains spanning enterprise analytics and an external benchmark (Table~\ref{tab:domains}). Three domains (finance, research portfolio, and supply chain) represent structured enterprise workflows over financial assets, internal R\&D projects, and supply chain entities. The fourth domain is the \emph{genius\_song\_lyrics} tool group from ToolBench, which provides an external API-based environment for music search and metadata retrieval with a distinct entity model and tool ecosystem derived from the ToolBench benchmark ~\cite{qin2023toolllm,guo2024stabletoolbench}.

\begin{table*}[t]
\centering
\small
\setlength{\tabcolsep}{4pt}
\begin{tabular}{@{}p{0.12\textwidth}p{0.16\textwidth}p{0.30\textwidth}p{0.30\textwidth}@{}}
\toprule
Domain & Target Entities & Representative Tools & Operational Focus \\
\midrule
Finance & Tickers, financial metrics & Symbol matching, statement retrieval, equity screening & Comparative financial analysis \\
Research portfolio & Projects, domains, dependencies & Project matching, portfolio filtering, readiness ranking & R\&D portfolio planning and prioritization \\
Supply Chain & SKUs, inventory, suppliers & Product matching, demand forecasting, bottleneck diagnosis & Logistics and demand planning \\
Music & Songs, artists, albums & Content search, detail retrieval, chart ranking & Music search \\
\bottomrule
\end{tabular}
\caption{Operational domains and representative tool sets used to evaluate SyntheticAgentTraceQA. The evaluation spans four domains: three enterprise analytics environments (finance, research portfolio, and supply chain) and the \emph{genius\_song\_lyrics} tool group from ToolBench. For each domain, the table summarizes the target entities, representative tools, and primary operational focus.}
\label{tab:domains}
\end{table*}
\paragraph{Shared operational abstraction.}
All domains share the operational taxonomy introduced in Section~\ref{taxonomy}. During profiling, each tool is assigned to an operational class and its parameters are annotated according to their functional roles. The resulting metadata are consumed by the same template generation, instantiation, execution, validation, and task-synthesis stages, independent of the underlying application domain.
\paragraph{External benchmark integration.}
To evaluate portability beyond our enterprise domains, we incorporated the \emph{genius\_song\_lyrics} tool group from ToolBench. Integration required only lightweight interface wrappers that normalize entity identifiers and parameter formats before profiling. No modifications were made to template generation, trace instantiation, execution, or validation, demonstrating that the framework transfers to external tool ecosystems through interface adaptation alone.

\section{Experiments}
\label{sec:experiments}

We investigate whether execution-grounded synthetic traces improve tool-augmented agents under realistic settings involving multi-domain tool use, execution feedback, and varying task difficulty.
\subsection{Hypotheses and Ablation Axes}

Our experimental design is structured around four hypotheses:

\begin{itemize}
    \item \textbf{H1 (Data utility):} Supervised fine-tuning on execution-validated traces improves tool-use reliability, Reference-Trace Agreement, and final-answer quality relative to the corresponding base models.
    
    \item \textbf{H2 (Reasoning initialization):} At a fixed model size, Thinking base models outperform their No-Thinking counterparts on execution-grounded reasoning tasks.

    \item \textbf{H3 (Process vs.\ outcome supervision):} For Thinking models, Masked supervision, which excludes \texttt{<think>} tokens from the training loss, leads to different performance than Full supervision, which computes the training loss over the complete assistant output.

   \item \textbf{H4 (Scale interaction):} Model size interacts with execution-grounded supervision, with larger models benefiting differently across evaluation metrics.
\end{itemize}
Our study considers the 4B and 9B variants of Qwen3.5. We first benchmark the pretrained checkpoints with and without Thinking enabled. We then fine-tune the models using two supervision strategies:

\begin{itemize}
    \item \textbf{Masked}: Excludes tokens within \texttt{<think>} blocks from the training loss while retaining supervision over executable tool calls and final-answer tokens.
    \item \textbf{Full}: Computes the training loss over both the reasoning trace and the final answer.
\end{itemize}
\subsection{Dataset Construction and Splits}

\paragraph{Generation protocol.}
Tool metadata (Profiler) were generated using \textit{Claude Sonnet 4.6}. Abstract templates were generated with eight LLMs: Claude Sonnet 4.6, Claude Sonnet 4.5, Amazon Nova Pro, Amazon Nova Lite, Qwen3-235B-A22B, Gemini 2.5 Pro, Claude Opus 4.1, and GPT-5.5. For each domain, templates were generated in batches of up to 30 instances (10 per difficulty level) with a temperature of 0.7. Generating templates independently for each domain reduces cross-domain contamination. The same LLM that generates a template is subsequently used for trace validation and task synthesis.

Grounded depth-first search (DFS) expanded 676 abstract templates (287 easy, 282 medium, and 107 hard) into 2,587 valid execution traces and 616 invalid traces, corresponding to a valid-trace rate of 80.8\%.

\paragraph{Task synthesis.}
Each validated trace is converted into three natural-language tasks (Basic, Intermediate, and Challenging), yielding 7,761 samples evenly distributed across difficulty levels. To preserve stylistic diversity, each task is generated by the same LLM that produced its source template, while the use of multiple generators prevents any single model from dominating the linguistic distribution. The annotator also generates a textual rationale describing how the recorded tool calls contribute to solving the task and producing the reference answer.
\paragraph{Dataset splits.}
For computational efficiency, we fine-tune on a subset of the generated corpus. The data are partitioned at the execution-trace level using a fixed random seed and stratified by trace complexity (easy, medium, and hard). All questions derived from the same validated execution trace are assigned to the same split, preventing data leakage. Because each execution trace generates one question at each difficulty level, stratifying by trace complexity while keeping traces intact also preserves a balanced distribution of question difficulties across the training, validation, and test sets. All reported results are evaluated on the held-out test set ($n=200$) and reported as mean $\pm$ standard error (SE) across tasks.

\subsection{Training Protocol}
To assess the quality and usefulness of the generated synthetic traces, we fine-tuned the target models on the training split of our dataset and evaluated their performance on the held-out test set. Our objective is not only to measure whether synthetic data can improve tool-use performance but also to understand how different supervision strategies over the reasoning process affect learning.

All fine-tuning runs share the same optimization setup to ensure that observed differences are attributable solely to the supervision strategy rather than to changes in hyperparameters. Specifically, we use LoRA adapters \cite{hu2021lora} with rank $r=16$ and scaling factor $\alpha=32$, train for three epochs using a learning rate of $2\times10^{-4}$, and quantize the base model to 4-bit NF4 precision. Due to computational resource constraints, training was performed on a fixed 1,000-example training split. Following standard instruction-tuning practice, the loss is computed only over assistant-generated tokens.

To isolate the effect of reasoning supervision (H3), we compare two training objectives. \emph{Full} supervision computes the loss over the entire assistant output, including the \texttt{<think>} blocks, whereas \emph{Masked} supervision excludes the reasoning blocks and computes the loss only over tool calls and final answers. All other training settings are kept identical.
\subsection{Dynamic Evaluation Protocol}\label{sec:evaluation}
Evaluating tool-augmented language models requires more than comparing final answers, as agents may recover from failed tool calls, revise their plans, or reach the correct solution through alternative execution trajectories. We therefore evaluate all models in a \textbf{multi-turn execution sandbox} that executes generated tool calls and feeds the resulting observations back to the model.

Each evaluation episode begins with a system prompt and a benchmark task. Models interact with the environment for up to eight turns using greedy decoding (temperature~0). Whenever a valid tool call is generated, the corresponding tool is executed and its output is appended to the conversation before the next generation step, enabling iterative planning and error recovery.

To match our evaluation semantics, tool calls generated inside \texttt{<think>} blocks are never executed. Only actions emitted after the closing \texttt{</think>} tag are interpreted as executable API calls.

The sandbox is fault tolerant: malformed arguments, invalid tool names, or schema violations return structured error messages instead of terminating the episode, allowing the model to recover from execution errors in subsequent turns.
\subsection{Metric Groups}

Following prior work on LLM-agent evaluation~\cite{yehudai2026survey}, we evaluate each model using three complementary groups of metrics: answer quality, tool behavior, and Reference-Trace Agreement.

\paragraph{Answer quality.}
\begin{itemize}
    \item \textbf{Answer-completion rate}: Percentage of tasks for which the agent produces a final answer.
    \item \textbf{Token-level F1}: Macro-averaged token-overlap F1 between generated and reference answers across all the tasks. Answers are lowercased and tokenized by whitespace before computing set-based precision, recall, and F1.
    \item \textbf{Numeric Match}: Macro-averaged fraction of reference numerical values correctly predicted across all the tasks. Numbers are extracted in order of appearance and matched positionally using an absolute tolerance of $0.01$.
\end{itemize}

\paragraph{Tool behavior.}
\begin{itemize}
    \item \textbf{Tool Attempt}: Average number of tool-call attempts per task.
    \item \textbf{Tool Success}: Average number of successfully executed tool calls per task.
    \item \textbf{Attempt-Success Gap ($\Delta$)}:
    \[
    \Delta = \frac{\text{Tool Attempt} - \text{Tool Success}}{\text{Tool Attempt}} \times 100.
    \]
    \item \textbf{Hallucinated Tool Rate}: Fraction of tool-call attempts that invoke an unregistered tool.
    \item \textbf{Syntax Error Rate}: Fraction of tool-call attempts rejected because of malformed tool-call syntax or invalid arguments.
\end{itemize}

\paragraph{Reference-Trace Agreement}
\begin{itemize}
    \item \textbf{Tool-set F1}: F1 score between the predicted and reference sets of tool calls.
    \item \textbf{Sequence Similarity}: Longest-common-subsequence (LCS) similarity normalized by the longer of the predicted and reference tool sequences.
    \item \textbf{Prefix Match}: Harmonic-weighted fraction of the reference prefix matched before the first deviation.
\end{itemize}

\section{Experimental Results}

Following the evaluation protocol described in Section~\ref{sec:evaluation}, we report results in three complementary categories: \emph{answer quality}, \emph{tool behavior}, and \emph{Reference-Trace Agreement}. We compare the base Qwen3.5-4B model with and without Thinking enabled, the larger Qwen3.5-9B model under the same inference settings, and two execution-grounded fine-tuned variants: \emph{Full}, which computes the training loss over the entire assistant output, including the reasoning trace, executable tool calls, and final-answer tokens, and \emph{Masked}, which excludes tokens within \texttt{<think>} blocks from the training loss while retaining supervision over executable tool calls and final-answer tokens.
\subsection{Answer Quality}
\label{sec:answer_quality}

Table~\ref{tab:answer_quality} reports final-answer performance in terms of Answer Completion, Token F1, and Numeric Match.

\paragraph{Effect of reasoning at inference.}

Enabling reasoning consistently improves all three answer metrics before fine-tuning. For the 4B model, Answer Completion increases from 21.0\% to 41.0\%, while Token F1 rises from 0.04 to 0.07 and Numeric Match from 2.09\% to 5.51\%. Similar improvements are observed for the 9B model, where Answer Completion increases from 25.0\% to 53.5\%, accompanied by higher Token F1 and Numeric Match. These improvements coincide with higher Reference-Trace Agreement and more successful tool use.

\paragraph{Effect of execution-grounded supervision.}

Execution-grounded fine-tuning further improves all three answer metrics. FT-4B-Masked increases Answer Completion to 59.5\% while achieving the highest Token F1 (0.15) and Numeric Match (9.83\%) among the 4B models. Likewise, FT-9B-Masked reaches the highest Answer Completion overall (86.0\%) while also improving Token F1 and Numeric Match relative to the corresponding reasoning baseline. These results support \textbf{H1}, showing measured improvements in answer production and agreement with the reference answers following execution-grounded supervision. Nevertheless, Token F1 and Numeric Match remain relatively low across all models, indicating substantial room for improvement in matching the reference answers.

\paragraph{Process supervision and model scale.}

Within both model sizes, \emph{Masked} supervision consistently outperforms \emph{Full} supervision across the reported metrics. For the 9B model, the Full variant performs below the Thinking baseline in Answer Completion and Token F1. Together, these results suggest that supervising executable tool calls and final answers, without optimizing over \texttt{<think>} tokens, is more effective than supervising the complete reasoning trace, partially supporting \textbf{H3}.

The effect of model scale is mixed. FT-9B-Masked achieves the highest Answer Completion (86.0\%), whereas FT-4B-Masked attains slightly higher Token F1 (0.15 vs.\ 0.14) and Numeric Match (9.83\% vs.\ 8.25\%). Thus, larger model size primarily improves answer production, while answer agreement with the reference remains comparable between the two masked models, partially supporting \textbf{H4}.

\begin{table}[t]
\centering
\caption{Final answer quality on the test set. Ans Compl. denotes the percentage of tasks with generated answers, Token F1 measures token-level overlap with reference answers, and Numeric Match measures the percentage of correctly predicted numerical values. Values are reported as mean $\pm$ standard error (SE), averaged over all test tasks. Higher is better ($\uparrow$).}
\label{tab:answer_quality}
\resizebox{\linewidth}{!}{
\begin{tabular}{lccc}
\toprule
Model &
Ans Compl. (\%) $\uparrow$ &
Token F1 $\uparrow$ &
Numeric Match (\%) $\uparrow$ \\
\midrule

Base-4B &
$21.00\pm2.89$ &
$0.04\pm0.01$ &
$2.09\pm0.81$ \\

Base-4B-Thinking &
$41.00\pm3.49$ &
$0.07\pm0.01$ &
$5.51\pm1.39$ \\

FT-4B-Masked &
$\mathbf{59.50\pm3.48}$ &
$\mathbf{0.15\pm0.01}$ &
$\mathbf{9.83\pm1.78}$ \\

FT-4B-Full &
$35.00\pm3.38$ &
$0.10\pm0.01$ &
$5.90\pm1.48$ \\
\midrule

Base-9B &
$25.00\pm3.07$ &
$0.04\pm0.01$ &
$2.19\pm0.82$ \\

Base-9B-Thinking &
$53.50\pm3.97$ &
$0.09\pm0.01$ &
$5.85\pm1.60$ \\

FT-9B-Masked &
$\mathbf{86.00\pm2.46}$ &
$\mathbf{0.14\pm0.01}$ &
$\mathbf{8.25\pm1.66}$ \\

FT-9B-Full &
$17.50\pm2.69$ &
$0.05\pm0.01$ &
$6.35\pm1.61$ \\
\bottomrule
\end{tabular}}
\end{table}

\subsection{Tool Behavior}
\label{sec:tool_behavior}

Table~\ref{tab:tool_behavior} summarizes tool-execution behavior, reporting the average number of attempted and successful tool calls together with the Attempt-Success Gap, Hallucinated Tool Rate, and Syntax Error Rate.

\paragraph{Effect of reasoning at inference.}

Reasoning substantially improves tool-use behavior before fine-tuning. Although the non-thinking models attempt more tool calls, they also exhibit substantially higher Attempt-Success Gaps and Hallucinated Tool Rates. Enabling reasoning reduces the average number of tool calls while increasing the proportion of successful executions. For example, the Attempt-Success Gap decreases from 18.9\% to 6.1\% for the 4B model and from 19.6\% to 0.07\% for the 9B model, with similar reductions in Hallucinated Tool Rate. These improvements coincide with higher Answer Completion and Reference-Trace Agreement.

\paragraph{Effect of execution-grounded supervision.}

Execution-grounded fine-tuning further improves tool behavior. Both Masked models reduce the Attempt-Success Gap, Hallucinated Tool Rate, and Syntax Error Rate while requiring fewer tool calls than their corresponding reasoning baselines. FT-4B-Masked decreases the Attempt-Success Gap from 6.1\% to 0.23\%, while FT-9B-Masked maintains the low error rates of Base-9B-Thinking using fewer tool calls on average (4.04 vs.\ 5.52). These results support \textbf{H1}, showing measured improvements in tool execution following execution-grounded supervision.

\paragraph{Process supervision and model scale.}

For both model sizes, the Masked and Full supervision strategies achieve low execution-error rates, with the Masked variants obtaining slightly lower Attempt-Success Gaps and Hallucinated Tool Rates. This observation is consistent with the answer-quality results in Section~\ref{sec:answer_quality}, where Masked supervision also produced stronger performance.

The effect of model scale is limited. Among the masked models, the 9B model performs fewer tool calls on average (4.04 vs.\ 5.33) while maintaining similarly low execution-error rates. Notably, the low Attempt-Success Gap observed for FT-9B-Masked is already present in the Base-9B-Thinking model, indicating that most of the improvement in execution reliability for the 9B model arises from inference-time reasoning, whereas execution-grounded supervision primarily reduces the number of tool calls. These observations partially support \textbf{H4}.
\begin{table}[t]
\centering
\caption{Tool execution performance on the test set. Attempt and Success denote the average number of attempted and successfully executed tool calls per query. $\Delta$ represents attempt–success gap, while Halluc. and Syntax denote invalid tool calls and malformed tool calls, respectively. Values are mean $\pm$ (SE) across test tasks; lower is better ($\downarrow$) for error metrics.}
\label{tab:tool_behavior}
\resizebox{\linewidth}{!}{
\begin{tabular}{lcc|ccc}
\toprule
Model &
Attempt &
Success &
$\Delta$ (\%) $\downarrow$ &
Halluc. (\%) $\downarrow$ &
Syntax (\%) $\downarrow$ \\
\midrule

Base-4B &
$7.11\pm0.14$ &
$5.64\pm0.23$ &
$18.87\pm2.72$ &
$18.05\pm2.69$ &
$0.45\pm0.38$ \\

Base-4B-Thinking &
$6.37\pm0.16$ &
$5.99\pm0.17$ &
$6.11\pm1.12$ &
$3.19\pm0.72$ &
$2.58\pm0.88$ \\

FT-4B-Masked &
$5.33\pm0.18$ &
$5.32\pm0.18$ &
$\mathbf{0.23\pm0.16}$ &
$\mathbf{0.23\pm0.16}$ &
$\mathbf{0.00}$ \\

FT-4B-Full &
$4.99\pm0.23$ &
$4.95\pm0.23$ &
$0.64\pm0.57$ &
$0.50\pm0.50$ &
$\mathbf{0.00}$ \\

\midrule

Base-9B &
$6.80\pm0.16$ &
$5.31\pm0.24$ &
$19.60\pm2.76$ &
$17.25\pm2.63$ &
$1.56\pm0.83$ \\

Base-9B-Thinking &
$5.52\pm0.21$ &
$5.51\pm0.21$ &
$\mathbf{0.07\pm0.06}$ &
$0.08\pm0.08$ &
$\mathbf{0.00}$ \\

FT-9B-Masked &
$4.04\pm0.15$ &
$4.04\pm0.15$ &
$\mathbf{0.07\pm0.07}$ &
$\mathbf{0.07\pm0.07}$ &
$\mathbf{0.00}$ \\

FT-9B-Full &
$6.68\pm0.18$ &
$6.68\pm0.18$ &
$0.05\pm0.05$ &
$0.06\pm0.06$ &
$0.00$\\
\bottomrule
\end{tabular}}
\end{table}
\subsection{Reference-Trace Agreement}
\label{sec:planning_quality}

Reference-Trace Agreement is evaluated independently of the final answer because multiple execution traces may correctly solve the same task. Table~\ref{tab:planning_quality} reports structural agreement between the predicted execution trace and the reference trace using Tool-set F1, Sequence Similarity, and Prefix.

\paragraph{Effect of reasoning at inference.}

Reasoning consistently improves Reference-Trace Agreement before any fine-tuning. Across both model sizes, the thinking variants achieve higher Tool-set F1, Sequence Similarity, and Prefix than their corresponding base models. These improvements indicate that explicit reasoning produces execution traces that more closely match the reference traces. In particular, the higher Sequence Similarity indicates closer agreement with the reference ordering of tool invocations. These findings support \textbf{H2}, showing that reasoning at inference provides a stronger foundation for execution-grounded reasoning by producing execution traces that more closely align with the reference traces.

\paragraph{Effect of execution-grounded supervision.}
Execution-grounded fine-tuning further improves Reference-Trace Agreement. The masked models achieve the highest Tool-set F1 (0.57 for both model sizes) while substantially improving Sequence Similarity relative to the corresponding reasoning baselines. These results support \textbf{H1}, showing measured improvements in agreement with the reference execution traces following execution-grounded supervision.

\paragraph{Process supervision and model scale.}

For the 4B models, both supervision strategies improve Reference-Trace Agreement relative to the reasoning baseline, although the Masked variant achieves higher Tool-set F1 and Prefix. The Full variant matches the Masked model in Sequence Similarity while performing worse on the remaining metrics, suggesting that supervising reasoning traces does not consistently improve agreement with the reference trace. This observation is consistent with the answer-quality results in Section~\ref{sec:answer_quality}, providing further evidence for \textbf{H3}.

The effect of model scale is mixed. FT-9B-Masked achieves higher Sequence Similarity (0.38 vs.\ 0.34), whereas FT-4B-Masked attains a slightly higher Prefix score (0.43 vs.\ 0.42), and both models obtain the same Tool-set F1 (0.57). Thus, larger model size primarily improves agreement with the reference execution order, while other Reference-Trace Agreement metrics remain comparable between the masked models, partially supporting \textbf{H4}.
\begin{table}[t]
\centering
\caption{Reference-Trace Agreement on the test set. Tool-set F1 measures overlap between the predicted and reference tool sets, Sequence Sim. measures normalized execution-order similarity, and Prefix measures agreement with the reference trace before the first deviation. Values are reported as mean $\pm$ standard error (SE) across test tasks; higher is better ($\uparrow$).}
\label{tab:planning_quality}
\resizebox{\linewidth}{!}{
\begin{tabular}{lccc}
\toprule
Model &
Tool-set F1 $\uparrow$ &
Sequence Sim. $\uparrow$ &
Prefix $\uparrow$ \\
\midrule

Base-4B &
$0.43\pm0.02$ &
$0.20\pm0.01$ &
$0.36\pm0.02$ \\

Base-4B-Thinking &
$0.53\pm0.02$ &
$0.27\pm0.01$ &
$0.36\pm0.02$ \\

FT-4B-Masked &
$\mathbf{0.57\pm0.02}$ &
$\mathbf{0.34\pm0.02}$ &
$\mathbf{0.43\pm0.02}$ \\

FT-4B-Full &
$0.54\pm0.02$ &
$\mathbf{0.34\pm0.02}$ &
$0.34\pm0.02$ \\
\midrule

Base-9B &
$0.46\pm0.02$ &
$0.22\pm0.01$ &
$0.32\pm0.02$ \\

Base-9B-Thinking &
$0.53\pm0.02$ &
$0.29\pm0.02$ &
$0.35\pm0.02$ \\

FT-9B-Masked &
$\mathbf{0.57\pm0.02}$ &
$\mathbf{0.38\pm0.02}$ &
$\mathbf{0.42\pm0.02}$ \\
FT-9B-Full &
$0.52\pm0.02$ &
$0.23\pm0.02$ &
$0.40\pm0.02$ \\
\bottomrule
\end{tabular}}
\end{table}

\subsection{Comparison with the Toucan Pipeline}

We compare \textsc{SyntheticAgentTraceQA} with the direct-generation pipeline of Toucan~\cite{xu2025toucan}. The generated datasets are evaluated using an LLM-as-a-Judge (Gemini 3.1 Flash Lite) with a rubric assessing question realism, clarity, naturalness, and the contribution of tool use to the final answer. As shown in Table~\ref{tab:llm_judge}, the judge assigns higher scores to \textsc{SyntheticAgentTraceQA} across all four criteria. In addition, \textsc{SyntheticAgentTraceQA} achieves a higher NovelSum score (0.50 vs.\ 0.39), indicating greater semantic diversity among the generated user queries.

Our reverse-generation pipeline also produces valid samples more efficiently, increasing the valid-trace rate to 80.8\% compared with 15\% for Toucan. As a result, the average generation cost (\$0.05 vs.\ \$0.12) and end-to-end latency (30\,s vs.\ 114\,s) per accepted trace are substantially reduced.

Finally, we fine-tune a reasoning-enabled Qwen3.5-4B model on each pipeline's dataset and evaluate it on the corresponding benchmark. As shown in Table~\ref{tab:holdout_pipeline}, both datasets improve performance over their respective base models. \textsc{SyntheticAgentTraceQA} yields larger gains in Tool-set F1, Sequence Similarity, and Tool Hallucination Rate, whereas Toucan produces a larger improvement in Answer Completion Rate (+0.230 vs.\ +0.195). Because each model is evaluated on its own benchmark, these results should be interpreted as within-pipeline improvements rather than a controlled cross-pipeline comparison.

\begin{table}[t]
\centering
\caption{Quality and diversity evaluation of synthetic data generated by SyntheticAgentTraceQA and Toucan~\cite{xu2025toucan}. LLM-as-a-Judge scores are reported on a 1-10 scale, and NovelSum~\cite{yang2025measuring} measures semantic diversity of generated user queries. Higher values indicate better quality or diversity ($\uparrow$).}
\label{tab:llm_judge}
\small
\begin{tabular}{lcc}
\toprule
\textbf{Metric} & \textbf{SyntheticAgentTraceQA} & \textbf{Toucan} \\
\midrule
Question Realism $\uparrow$      & \textbf{9.85} & 9.35 \\
Question Clarity $\uparrow$      & \textbf{9.85} & 9.70 \\
Question Naturalness $\uparrow$  & \textbf{9.89} & 8.02 \\
Tools Improved Answer $\uparrow$ & \textbf{8.56} & 8.38 \\
\midrule
NovelSum $\uparrow$      & \textbf{0.50} & 0.39 \\
\bottomrule
\end{tabular}
\end{table}

\begin{table}[t]
\centering
\caption{Comparison of fine-tuning results using datasets generated by \textsc{SyntheticAgentTraceQA} and Toucan~\cite{xu2025toucan}. Values report Qwen3.5-4B performance after fine-tuning, with improvements over the corresponding base model shown in parentheses. Each model is evaluated on its respective benchmark; therefore, the reported scores should be interpreted as within-pipeline results rather than as a controlled cross-pipeline comparison. Higher is better ($\uparrow$) except for Tool Hallucination Rate ($\downarrow$).}
\label{tab:holdout_pipeline}
\small
\setlength{\tabcolsep}{3pt}
\begin{tabular}{lcc}
\toprule
\textbf{Metric} & \textbf{SyntheticAgentTraceQA} & \textbf{Toucan} \\
\midrule
Tool-set F1 $\uparrow$
& \textbf{0.612} (+0.086)
& 0.596 (+0.061) \\

Sequence Similarity $\uparrow$
& \textbf{0.492} (+0.117)
& 0.470 (+0.086) \\

Tool Hallucination Rate $\downarrow$
& \textbf{0.000} ($-0.003$)
& 0.002 (+0.001) \\

Answer Completion Rate $\uparrow$
& \textbf{0.795} (+0.195)
& 0.685 (+0.230) \\
\bottomrule
\end{tabular}
\end{table}
\section{Discussion}

\paragraph{Execution-grounded supervision improves tool-use behavior.}
Our experiments show that execution-grounded supervision improves tool behavior, Reference-Trace Agreement, and final-answer generation across the evaluated environments and model configurations. These gains suggest that validated execution traces provide supervision beyond task-answer pairs by exposing models to consistent tool choices, parameter dependencies, and intermediate outputs. This observation aligns with prior work showing that tool-use demonstrations and interaction trajectories can improve agent capabilities~\cite{qin2023toolllm,zhou-etal-2026-toolgrad,xu2025toucan}.

Our contribution is not execution-first generation alone, but the combination of operational tool and parameter profiling, active-domain validation, abstract workflow generation, dependency-aware tool assignment, and execution-grounded task synthesis. The comparison with Toucan should be interpreted cautiously because the two approaches use different evaluation benchmarks; however, the results indicate that verified executions are a useful source of supervision for tool-augmented agents.

\paragraph{Reasoning supervision introduces a process-outcome trade-off.}
Our results reveal a trade-off between supervising full assistant outputs and masking reasoning tokens. Masked supervision achieves stronger answer-generation performance across the evaluated settings, whereas Full supervision yields weaker answer completion and does not consistently improve reference-trace agreement—most notably for the 9B model, where Full supervision falls below the Thinking baseline on several metrics (e.g., Answer Completion drops to 17.5\%). A possible explanation is that the reasoning traces used for supervision are post-hoc rationales describing how recorded tool calls support the answer, rather than necessarily the model’s unique reasoning process. Prior work suggests that reasoning supervision depends on trace quality, teacher-student compatibility, and the training objective~\cite{huang2026fine,yang2026reasoning}.

\paragraph{Execution-first synthesis provides validated supervision data.}
\textsc{SyntheticAgentTraceQA} constructs and validates executable workflows before synthesizing user tasks and answers, reducing invalid tool interactions during data generation. Similar trajectory-based approaches have highlighted the value of execution-grounded supervision for tool-augmented agents~\cite{armengol2025execute,wang2026trajectory2task}. Our dataset analysis further shows higher judge-assessed realism, clarity, and naturalness scores from an LLM evaluator, as well as higher NovelSum semantic diversity~\cite{yang2025measuring}. These results reflect automated evaluation measures rather than human-validated quality.

\paragraph{Model scale effects are metric-dependent.}
Increasing model size does not uniformly improve all metrics. FT-9B-Masked achieves the highest answer-completion rate and sequence similarity, while FT-4B-Masked obtains slightly higher Token F1, Numeric Match, and Prefix agreement. Tool-set F1 is comparable between scales, suggesting that model capacity benefits specific aspects of agent behavior rather than providing uniform improvements.

\paragraph{Limitations and future directions.}
Our evaluation focuses on controlled tool ecosystems with fixed schemas and execution. Due to computational constraints, fine-tuning used only a subset of the generated dataset. Future work will study scaling across different training set sizes, dynamic environments with changing APIs and longer-horizon workflows, and cross-domain tool composition.
\section{Conclusion}

We introduced \textsc{SyntheticAgentTraceQA}, an execution-first framework for generating supervision data for tool-augmented agents. Rather than generating tasks and inferring tool trajectories afterward, our framework first constructs and validates executable tool-use traces, then synthesizes user tasks, teacher-generated reasoning annotations, and reference answers from successful executions. This design enables controlled generation of tool-use examples with validated dependencies and executable workflows.

Experiments across four tool ecosystems and multiple model configurations show that training on execution-grounded traces improves tool behavior, Reference-Trace Agreement, and final-answer generation on the evaluated benchmarks. Our analysis further reveals a trade-off between supervision strategies: Masked supervision, which excludes <think> tokens while retaining tool-call and answer supervision, achieves stronger answer-quality metrics, whereas Full supervision over the complete assistant output yields weaker answer completion and does not consistently improve reference-trace agreement—most notably for the 9B model, where Full supervision falls below the Thinking baseline on several metrics. These results suggest that the optimal supervision strategy depends on the target capability being optimized.

\bibliographystyle{ACM-Reference-Format}
\bibliography{latell}

\end{document}